# Y Chromosome of Aisin Gioro, the Imperial House of Qing Dynasty

YAN Shi[1]*, TACHIBANA Harumasa[2], WEI Lan-Hai[1], YU Ge[1], WEN Shao-Qing[1], WANG Chuan-Chao[1]

[1] Ministry of Education Key Laboratory of Contemporary Anthropology and State Key Laboratory of Genetic Engineering, Collaborative Innovation Center for Genetics and Development, School of Life Sciences, Fudan University, Shanghai 200438, China

[2] Pen name

* Please contact yanshi@fudan.edu.cn

## Abstract

House of Aisin Gioro is the imperial family of the last dynasty in Chinese history – Qing Dynasty (1644 – 1911). Aisin Gioro family originated from Jurchen tribes and developed the Manchu people before they conquered China. By investigating the Y chromosomal short tandem repeats (STRs) of 7 modern male individuals who claim belonging to Aisin Gioro family (in which 3 have full records of pedigree), we found that 3 of them (in which 2 keep full pedigree, whose most recent common ancestor is Nurgaci) shows very close relationship (1 – 2 steps of difference in 17 STR) and the haplotype is rare. We therefore conclude that this haplotype is the Y chromosome of the House of Aisin Gioro. Further tests of single nucleotide polymorphisms (SNPs) indicates that they belong to Haplogroup C3b2b1*-M401(xF5483), although their Y-STR results are distant to the "star cluster", which also belongs to the same haplogroup. This study forms the base for the pedigree research of the imperial family of Qing Dynasty by means of genetics.

This research was supported by the grants from the National Science Foundation of China (31271338), and from Ministry of Education (311016).

# 爱新觉罗皇族的Y染色体

严实[1]* 橘玄雅[2] 韦兰海[1] 余歌[1] 文少卿[1] 王传超[1]

[1] 复旦大学生命科学院，现代人类学教育部重点实验室、遗传工程国家重点实验室、遗传与发育协同创新中心，中国，上海 200438

[2] 笔名

* 请联系 yanshi@fudan.edu.cn

## 摘要

爱新觉罗氏为中国最后一个朝代——清朝（1644-1911）的皇族。本研究通过对7个自称属于爱新觉罗的现代家族（其中3家保留有完整家谱）男性的Y染色体短串联重复片段（STR）分析，发现其中3家（其中2家有完整家谱，其最近共祖为清太祖努尔哈赤）的Y染色体17 STR相互间只相差1–2步突变，且极为罕见，因此确定为爱新觉罗家族的类型。对此3个样本的单核苷酸多态位点（SNP）的测试结果表明此类型属于C3b2b1*-M401(xF5483)单倍群，尽管此3个样本的STR组合与同属该单倍群的"星簇"相距很远。本研究为今后的女真、满族源流及清朝皇族家族史的研究提供了重要参照。

关键词：Y染色体，父系，族谱，家族史，单倍群，清朝

The non-recombining region of male-specific Y chromosome is strictly passed on from father to son. Random mutations may happen during the inheritance, therefore we can reconstruct the paternal phylogeny by comparing the sequence of males' Y chromosomes. The males sharing nearer common ancestor have more similar sequences, and *vice versa*. Commonly used markers for tracing Y chromosome phylogeny include single nucleotide polymorphism (SNP) and short tandem repeat (STR). The SNPs that have a higher frequency in a population usually can be used for tracing relatively old (usually more than thousands of years) clades. On the other hand, using high-throughput sequencing of long regions of Y chromosome, we can discover rare and recent mutations, and calculate the divergence times by counting the number of SNPs, with a theoretical highest precision of ~100 years, i.e. 3 – 4 generations[1]. The mutation rate of STRs is higher, and STR recurrent mutations occur frequently, therefore relatively late relationships can be traced by comparing STRs[2].

Since the surname inheritance as well as genealogical records in most East Asian populations follow the paternal lineage, just like the Y chromosome, we can verify the precision of genealogy records using the Y chromosome. In the following cases, different types of Y chromosomes occur in one paternal family: illegitimate son, adoption from another family, and when a son inherited his mother's surname. Adoption from inside a paternal family usually cannot be recognized through Y chromosome. Also, it commonly happens that a family traces their genealogy to an irrelative family with the same surname, in order to enhance their own prestige.

Due to the uncertainty of father-son relationship, technically, the Y chromosomal result of a man shows only the haplotype of himself rather than that of his whole paternal family. However, corresponding results of two modern individuals that were supposed to be akin would verify the lineages from their most recent common ancestor to the both individuals, and the more tested results would make the conclusion more reliable. When the lineage is distant (more than 600 years, before the migration wave in early Ming Dynasty), the uncertainty of parentage accumulates, and it is common that each descendant family shows a different result. In that case, the result of ancient DNA would be more reliable than that concluded through modern individuals' results and genealogy records. But attentions must be paid that the identity of an ancient sample could also be problematic. And due to the DNA degeneration over time[3], the results of ancient DNA can be far less precise than modern samples due to unsuccessful DNA extraction and amplification, and for many cases the results may be not retrievable at all.

Currently, the only reported Y chromosome result of historical Chinese imperial houses is that of Cáo Wèi Dynasty (220 – 265 AD), belonging to the Haplogroup O2*-M268. This conclusion was first achieved via comparison and statistics of the Y chromosomes of descendants recorded in the genealogy books, then verified using the DNA extracted from the remains of Cáo Dǐng (曹鼎), Cáo Cào (曹操)'s grandfather's brother[4, 5]. The Y chromosome type of Genghis Khan was once supposed to belong to the "star cluster" under C3*-M217(xM48)[6], however, this was only deduced based on the Mongolian STR cluster with high frequency without genealogical support, and still under debate[7].

By comparing the relics of the last Russian Tsar – Nicholas II, and one of his extant nephew, the Y chromosome type of Romanov Dynasty was concluded as Haplogroup R1b[8]. The House of Bourbon was determined to be member of Haplogroup R1b[9] from three Y chromosomes of present day, which overturned a former result of Haplogroup G, deduced from a handkerchief soaked with Louis XVI's blood. Y chromosome of Napoléon Bonaparte was proved to be member of Haplogroup E1b1b1c1-M34 by testing his beard hair's basis and sample from one of his present-day relatives[10]. In addition, the Y chromosome types of Lithuanian royal family and some ancient Egyptian dynasties have also been tested.

The House of Aisin Gioro (愛新覺羅, all the Manchurian names in this article apply to the Latin transcription invented by P. G. von Möllendorff[11]) originated from Jurchen tribes, and finally evolved into Manju (commonly Manchu) and founded the last imperial dynasty in China – the Qing Dynasty (1644 – 1911). There has been only one study speculating the Y chromosomal type of the House of Aisin Gioro: Xue *et al.*[12] revealed an STR cluster under Haplogroup C3c-M48 includes ~ 3.3% of males in some populations Northeast China as well as in Mongolians, and named it as the "Manchu cluster", and calculated its divergence time at 590 ± 340 years. The authors claimed that due to its recency, the expansion should be an event recorded in history, and due to its large population, the only explanation should be the Y chromosome of Aisin Gioro, and the most recent common ancestor is Giocangga (覺昌安), the grandfather of Nurgaci (努爾哈赤, commonly Nurhaci, 1559 – 1626, the first emperor of Later Jīn Dynasty). However, the authors did not pose any genealogical evidences. Besides, the figure of this study showed that the "Manchu cluster" has clearly higher frequency in Oroqen, Ewenki, Hezhe, Inner and Outer Mongolia than in Manchu and Sibe. We think the conclusion of this study need to be verified by further genealogical studies.

In Manchurian language, the term "*hala*" (哈拉) is the surname of people who are said to have the same common ancestor in paternal lineage, but it was also suggested that the people in one *hala* only belonged to the same tribe rather than having the same paternal origin[13]. The Gioro *hala* (覺羅哈拉) includes those sub-surnames like Aisin Gioro, Irgen Gioro (依爾根覺羅), Šušu Gioro (舒舒覺羅) *etc.*, and "*aisin*" means "gold" in Manchurian. Aisin Gioro includes the people with the same traceable ancestor with Nurgaci, *i.e.* the descendants of his great-grandfather Fuman (福滿). Since 1636, the descendants of Taksi (塔克世, father of Nurgaci) were nominated as "*uksun*" (宗室, "the imperial house"), while the peripheral members in Aisin Gioro as "*gioro*" (覺羅), which has a different definition to the Gioro *hala*. After conquering the whole China, the descendants of Kangxi Emperor (1654 – 1722) were called "near *uksun*" (近支宗室) and the others "far *uksun*" (遠支宗室).

In a statistics in 1915, there are 27,884 males of Aisin Gioro alive, including 16,454 *uksun* and 11,430 *gioro*[14]. Considering natural population growth, as well as wars, political events, and the policy of family planning in the recent 100 years, currently estimated number of Aisin Gioro males is less than 30,000. Due to a decree in Qing Dynasty that *uksun* were forbidden to leave the capitals – Beijing and Mukden (presently Shenyang), great number of present Aisin Gioro still distribute in these two cities.

## Materials and Methods

We collected blood or oral samples from 7 males who lived for generations in Beijing or Liaoning, who all claimed to belong to the House of Aisin Gioro but diverged more than five generations with each other. The study was under the approval of the Ethics Committee of Biological Research at Fudan University, and all the samples were collected with the Informed Consent. Among the 7 males, 3 have complete genealogical records inside the book of "Genealogy of Aisin Gioro"[15], while the others could only offer partial ancestral information. All the 3 with full lineage were "far *uksun*", and lived in Beijing for generations, 2 of them (Samples A and D) being descendants of Dodo (多鐸, the 15th son of Nurgaci), and the other (Sample B) from the House of Prince of Sù (肅親王), descendant of Hooge (豪格, the first son of the second Qing emperor Hung Taiji 皇太極). In addition, one testee from Beijing (Sample E) claims to be from the family of Prince of Zhèng (鄭親王), descendant of Jirgalang (濟爾哈朗), nephew of Nurgaci. A further testee (Sample C) from Beijing claim to be descendant of "*gioro*

Langkio" (覺羅郎球), who is possibly the great-grandson of Soocangga (索長阿), the third son of Fuman. The other two testees are from Liaoning Province, one from Jinzhou (錦州) (Sample F) also claimed descendant of Dodo, and the other (Sample G, without knowing his ancestors' names) from Benxi (本溪). In order to avoid troubles or loss of reputations of the testees and their close relatives, we do not reveal all the testees' names as well as the information of their near ancestry.

We extracted DNA of all these samples and amplified 17 Y-STRs (DYS19, DYS389I/II, DYS390, DYS391, DYS392, DYS393, DYS437, DYS438, DYS439, DYS448, DYS456, DYS458, DYS635, Y-GATA H4, DYS385a/b) using Y-Filer kit (Life Technologies, CA, USA). For those samples that 17-STRs fell into the range of Haplogroup C, we genotyped 7 following SNPs using Sanger sequencing: M217, F1396, F3535, F3273, F5483/SK1074, M546, and M401 to determine their fine phylogeny (see Table 1). Network of C3 samples from this study and compared samples were drawn using Network 4.6.1.2 (Fluxus) with Median Joining algorithm[16].

Table 1    Tested SNPs

| SNP | Position (hg19) | Mutation | Forward primer | Reverse primer |
| --- | --- | --- | --- | --- |
| M217 | 15437333 | A>C | ccagtatctccaaaatcctctcgta | cgtaagcatttgataaagctgctgt |
| F1396 | 8746611 | C>T | cacacccctcttgtgtcacagca | acctgcctgtgtgcagtctttcta |
| F3535 | 23768635 | G>T | tcaccaggttgtatcctcacctgt | agaaccacacagcctcatgagctg |
| F3273/M504 | 21888793 | T>C | cttttgataaaagtatcatgtgttcagt | cacaatgtagttgaaaattgaagttgta |
| M546 | 19060966 | C>A | ccttgtaaaagccaagctgccctt | ttaggttttacaatagtgatgctattct |
| M401 | 2745569 | TG ins | gctggtctcaaacttctgacctttg | gctggtctcaaacttctgacctttg |
| F5483/SK1074 | 19086774 | T>A | aaagtctgccattatgatgatatagttc | cttcaagtatacgtgaaggctaaatg |

## Results

Among the 17-STR results of all the 7 testees, 3 (A – a descendant of Dodo, B – descendant of Hooge, C – descendant of Langkio by oral history, see Fig. 1) showed similar results, with only one or two steps of Y-STR mutations among each other (Table 2, Fig. 2), while the others have at least 6 steps to them and among each other, which excludes the possibility that they share a common ancestor inside 500 years. By searching published literatures, we found two nearest samples that have two steps of difference with Sample A – one is a Manchu from Xinbin (新賓) County of Liaoning Province[17], the other is from northeastern part of Mongolia[18]. We also searched the largest Y-STR dataset, ysearch.org, but could not find any samples so close as 2 steps.

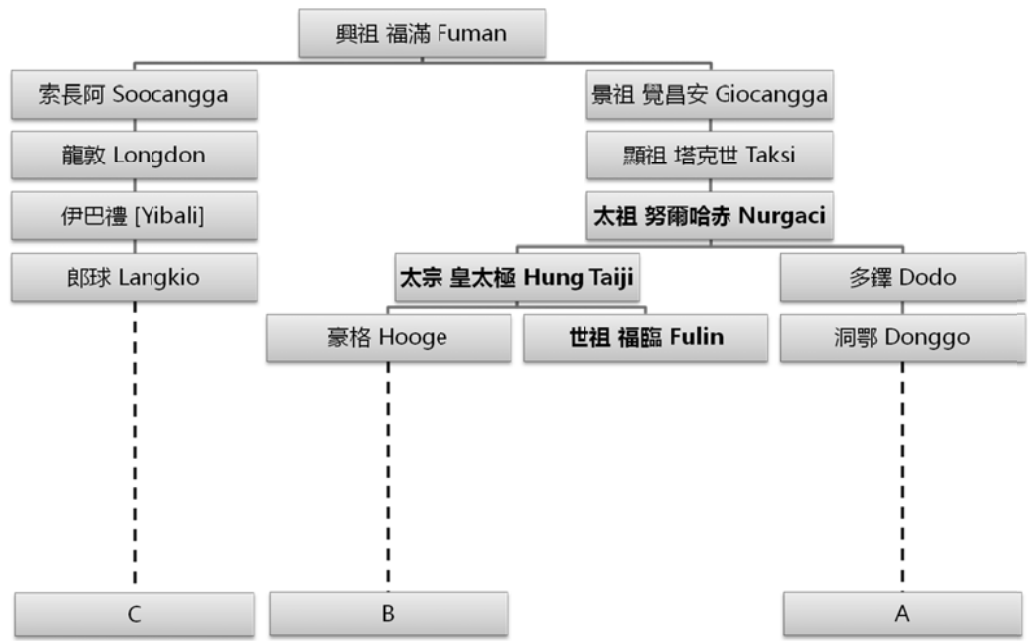

Fig. 1　The genealogy of the three samples (A, B, C) which have close STR results. The genealogy of A and B is fully described, while C has only oral history. Those labeled in bold are emperors of Later Jīn and Qīng Dynasty.

Table 2　The STR results of the tested samples and those for comparison (B as the central value, and those loci different to B are shaded gray).

| Individual | Clade | DYS 19 | DYS 389I | DYS 389b | DYS 390 | DYS 391 | DYS 392 | DYS 393 | DYS 437 | DYS 438 | DYS 439 | DYS 448 | DYS 456 | DYS 458 | DYS 635 | Y-GATA H4 | DYS 385a | DYS 385b |
|---|---|---|---|---|---|---|---|---|---|---|---|---|---|---|---|---|---|---|
| B (Hooge's) | C-M401 | 15 | 13 | 16 | 24 | 10 | 11 | 14 | 14 | 10 | 11 | 21 | 16 | 19 | 23 | 11 | 12 | 12 |
| A (Dodo's) | C-M401 | 15 | 13 | 16 | 24 | 10 | 11 | 14 | 14 | 10 | 11 | 21 | 15 | 19 | 23 | 11 | 12 | 12 |
| C (Langkio's) | C-M401 | 15 | 13 | 16 | 24 | 10 | 11 | 14 | 14 | 11 | 11 | 21 | 16 | 19 | 23 | 11 | 12 | 12 |
| D (Dodo's) | O-002611 | 17 | 12 | 17 | 24 | 10 | 13 | 12 | 14 | 10 | 12 | 18 | 13 | 17 | 21 | 13 | 12 | 18 |
| E (Jirgalang's) | C-M401 ("star cluster") | 15 | 13 | 16 | 25 | 10 | 11 | 13 | 14 | 10 | 10 | 22 | 15 | 19 | 21 | 11 | 12 | 12 |
| F (Jinzhou) | O-M176 | 15 | 14 | 16 | 22 | 10 | 13 | 13 | 14 | 13 | 11 | 18 | 17 | 18 | 21 | 12 | 10 | 20 |
| G (Benxi) | C-F1144 | 15 | 14 | 16 | 24 | 10 | 11 | 14 | 14 | 11 | 11 | 20 | 15 | 16 | 21 | 12 | 11 | 20 |
| Ht188 (Xinbin Manchu) |  | 15 | 13 | 16 | 24 | 10 | 11 | 14 | 14 | 10 | 11 | 21 | 15 | 18 | 24 | 11 | 12 | 12 |
| MG3_5 (Mongolian) |  | 15 | 13 | 16 | 24 | 10 | 11 | 14 | 14 | 10 | 11 | 22 | 15 | 19 | 24 | 11 | 12 | 12 |
| Central value of "star cluster" | C-M401 | 16 | 13 | 16 | 25 | 10 | 11 | 13 | 14 | 10 | 10 | 22 | 15 | 18 | 21 | 11 | 12 | 13 |
| Central value of "Manchu cluster" | C-M48 | 16 | 13 | 16 | 24 | 9 | 11 | 13 | 14 | 10 | 11 | 20 | 15 | 17 | 23 | 10 | 12 | 13 |

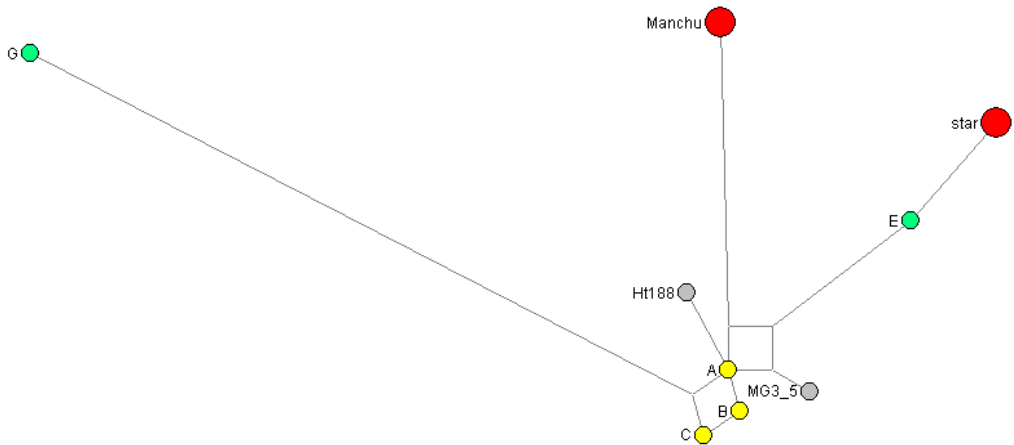

Fig. 2   Network of C3 samples from this study (yellow for concluded Aisin Gioro and green for two others, E and G), as well as two samples (Ht188 and MG3_5, in gray) and the center value of "star cluster" and "Manchu cluster") for comparison.

Among the 7 tested SNPs, M217 (Haplogroup C3) was well known in numerous literatures; F1396, F3535, and F3273/M504 were discovered from high-throughput sequencing of Y chromosomes of Mongolian and Manchu samples[19]; M401 and M546 were found in Mongolian and many related populations, especially the "star cluster", which was suspected to be the clan of Genghis Khan, also belongs to the clade C3b2b1-M401[18] (Fig. 3); and F5483/SK1074 from Orochon and Daur samples[20]. The samples A, B, and C are genotyped as M217+, F1396+, F3535+, F3273+, M546+, M401+, and F5483- ("-" for ancestral, and "+" for derived), i.e., belong to Haplogroup C3b2b1*-M401(xF5483), the same as the "star cluster", despite of 10 steps of difference in 17 STRs between Sample B and the central value of the "star cluster" (Table 2).

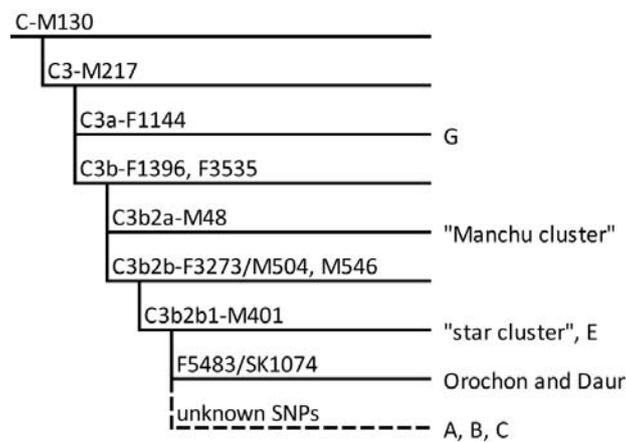

Fig. 3   Phylogenetic tree of Haplogroup C concerning the SNPs genotyped in this study

## Discussions

Although there is large confidence interval for the divergence time estimated using Y-STRs, and there is still debate about the STR mutation rate (from the "effective mutation rate" of $6.9\times10^{-4}$ in 25 years, *i.e.* one mutation per 36,231 years[21], to the father-to-son mutation rate of $3\times10^{-3}$ per generation[22], *i.e.* one mutation per 10,000 years), the calculated age can give a rough image about the divergence time. Using the software NETWORK 4.6.1.2 (Fluxus) with a medium mutation rate of one mutation of 20,000 years for 15 STRs with exceptions of DYS385a/b (since DYS385 loci may recombine with each other and cause complexity in time estimation), we estimated the most recent ancestor of A, B, and C as 666±471 years before present. This time is in accordance with the genealogy (Fuman as the most recent common ancestor, ~ 500 years before present). Adding the Xinbin Manchu sample Ht188, the calculated divergence time becomes 1333±653, longer than Aisin Gioro's history. Therefore, we suggest the Y-STR combination of the 3 samples be that of Aisin Gioro family.

Since this Y-STR haplotype is rare in the modern populations, including Han, Manchu and Mongol, and we found the samples nearer than 2 steps to Sample B are only from Aisin Gioro family, therefore when a member in question with less than 2 steps to this haplotype has a great chance that he really belongs to Aisin Gioro. Xinbin County in Liaoning is the location of Hetu Ala (赫圖阿拉) City where Nurgaci arose, and there can be certain amount of Aisin Gioro's relatives (with divergence time of 500 to thousands of years). However, we found only one sample (Ht188) in 231 published Manchu data that is 2 steps of STR difference to Sample A, which means that the Aisin Gioro family is not frequent in a normal Manchu population. However, we still expect to find more individual with this type in Manchu of Beijing and Shenyang. This study also shows that, among those who claim to be Aisin Gioro without a clear genealogy, the chance that one really belongs to Aisin Gioro (in biological sense) is not high.

Understanding of Y chromosomal type of Aisin Gioro will benefit the historical researches about Jurchen, Manchu and Qing Dynasty, and will help to reveal the truth for some rumors. For example, some scholars inside Aisin Gioro claims that Gioro *hala* were descendants of Huīzōng of Sòng Dynasty (宋徽宗, 1082 – 1135), who was captured by the Jurchen Jīn Dynasty and exiled to Northeast China. There is also rumor that Qianlong Emperor is not the biological son of Yongzheng Emperor; although no descendant of Qianlong was included in this study, it can be easily verified in the future.

It should be noted that during the Chinese history, the father-son relationship in social law is not equal to that in biological sense. Many adoption events due to sonlessness were recorded in the genealogy of Aisin Gioro; most of the adoption events were between closely related clans, however, adoption from another surname also happened. Therefore it is not proper to judge a social family heredity simply using the biological proofs.